\documentclass[sn-mathphys]{sn-jnl}

\jyear{2023}%

\def\sk{\hat{s}_k}
\def\kmax{k_{\rm max}}
\newcommand\bet{{g}}
 
\newcommand\alps{{\frac{\hbar^2}{2m}}}
\newcommand\mub{{\mu}}

\newcommand\dertt[1]{ \frac{\partial{ #1}}{\partial t} }
\newcommand\gd{\mbox{${\bf \nabla}^{2}$}}
\newcommand\psib{\overline{\psi}}

\begin{document}

\title[Kelvin waves, mutual friction, and fluctuations in the GP model]{Kelvin waves, mutual friction, and fluctuations in the Gross-Pitaevskii model}

\author*[1]{\fnm{Giorgio} \sur{Krstulovic}}\email{krstulovic@oca.eu}

\author[2]{\fnm{Marc E.} \sur{Brachet}}\email{brachet@phys.ens.fr}

\equalcont{These authors contributed equally to this work.}

\affil*[1]{Universit\'e C\^ote d'Azur, Observatoire de la C\^ote d'Azur, CNRS, Laboratoire Lagrange, Bd de l'Observatoire, CS 34229, 06304 Nice cedex 4, France}

\affil[2]{Laboratoire de Physique de l'Ecole Normale Sup\'erieure, ENS, Universit\'e PSL, CNRS, Sorbonne Universit\'e, Universit\'e de Paris,
F-75005 Paris, France}



\abstract{In this work we first briefly review some of the mutual friction effects on vortex lines and rings that were obtained in the context of the truncated Gross-Pitaevskii equation in references Krstulovic \& Brachet [Phys.~Rev.~E \textbf{83}(6), 066311 and Phys.~Rev.~B \textbf{83}132506 (2011)], with particular attention to the anomalous slowdown of rings produced by thermally excited Kelvin waves.
We then study the effect of mutual friction on the relaxation and fluctuations of Kelvin waves on straight vortex lines by comparing the results of full $3D$ direct simulations of the truncated Gross-Pitaevskii equation with a simple stochastic Local-Induction-Approximation model with mutual friction and thermal noise included. This new model allows us to determine the mutual friction coefficient $\alpha$ and $\alpha'$ for the truncated Gross-Pitaevskii equation. }




\maketitle


\section{Introduction}\label{sec:intro}

Our knowledge and understanding of the physics of superfluid helium and quantum turbulence has greatly progressed \cite{Vinen2002}. However, at present, there is no single theory that comprehensively describes all the systems and can predict the observed effects across all length and time scales. Therefore, much of the progress in this field relies on phenomenological models, which may be better suited for certain types of problems than others, and are of several types.
\begin{itemize}
\item The phenomenological two-fluid model, proposed independently by L.~Tisza and L.~Landau, considers the fluid as a mixture of two components: the superfluid and the normal fluid. It successfully describes many flow properties of superfluid helium, including the propagation of second sound. However, it does not account for the presence of quantized vortices, which are an important feature of quantum flows, and were experimentally discovered to have a circulation of $h/m$ by Vinen in 1961 \cite{Vinen1961}.
\item An extension of the two-fluid model is the Hall-Vinen-Bekharevich-Khalatnikov model \cite{HallVinen1_1956,HallVinen2_1956,Bekarevich61}, which includes the effect of interactions between quantized vortices and the normal fluid through a mutual friction term. However, this model ignores the distinction between individual vortices and only considers length scales larger than the mean separation between vortices, making it an effective, coarse-grained model suitable for describing superfluid turbulence at low Mach numbers.
\item The vortex filament model \cite{Schwarz_85,Schwarz_88} overcomes some of the limitations of the two-fluid and HVBK models by treating the quantised vortices as filaments in three dimensions and evolving them under the Biot-Savart law plus a mutual friction term that mimics the coupling between the normal and superfluid components. In this model, quantum vortex reconnection is implemented by an ad-hoc algorithm and the normal fluid is assumed as given and quantum vortices do not modify it.
\item The self-consistent vortex filament and Navier-Stokes models \cite{Kivotides_TripleVortexRing_2000a,Yui_ThreeDimensionalCoupledDynamics_2018,Galantucci_FrictionenhancedLifetimeBundled_2023,Galantucci_NewSelfconsistentApproach_2020}. This approach allows to account for the mutual interaction of vortices and the normal fluid.
\item Finally, at zero or near-zero temperatures, and for weakly interacting
bosons, the Gross-Pitaevskii (GP) equation provides a good hydrodynamical description of a quantum
flow, that naturally includes quantum vortices as exact solutions which
can reconnect.
\end{itemize}

A problem with the GP model is that including finite temperature effects can be difficult \cite{Gardiner00,Gardiner02,Calzetta07,Berloff14}. 
Note that, to the best of our knowledge, only a few attempts \cite{Coste1998,brachet2022coupling} have been made to self-consistently couple the Gross-Pitaevskii equation to a Navier-Stokes description of the normal fluid.

In the present study, a minimalistic approach is used by considering classical field models, which involve spectrally truncating the GP equation \cite{Proukakis08}. Long time integration of the truncated system results in microcanonical equilibrium states that can capture a condensation transition, which has been previously demonstrated \cite{Davis01}. The same transition was later reproduced using a grand-canonical method, where it was shown to be a standard second-order $\lambda$-transition \cite{Krstulovic11}. This approach also correctly captures dynamical counterflow effects on vortex motion, such as mutual friction and thermalisation dynamics, which have been investigated in previous studies \cite{Krstulovic11, Krstulovic11a}.

{This manuscript begins by offering a concise review of the mutual friction effects on vortex lines and rings, that were observed in our previous numerical studies. Subsequently, we present new, original results that are obtained by performing full 3D simulations of the truncated Gross-Pitaevskii equation. Focusing on the behavior of vortex lines, these results are then compared to those derived from a simplified stochastic Local-Induction-Approximation model. This paper} is structured as follows: In Section~\ref{sec:GP}, we provide a detailed explanation of the GP model and its numerical implementation. Specifically, in Section~\ref{sec:basic} we discuss the basic zero-temperature GP theory and its conserved energies. We also introduce and discuss the properties of Kelvin waves propagating along quantum vortices.
In Section~\ref{sec:trunc}, we review our spectral truncation methods for incorporating finite-temperature effects, which were developed in reference \cite{Krstulovic11}. We demonstrate the equivalence between the micro-canonical and canonical statistical ensembles in Section~\ref{sec:equiv}. 
Section~\ref{sec:mutual} reviews our previous findings on mutual friction in lines and rings, as presented in references \cite{Krstulovic11b} and \cite{Krstulovic11}. Specifically, standard effects are found for lines in Section~\ref{sec:lines}, while rings are shown to undergo an anomalous slowdown, produced by thermally excited fluctuating Kelvin waves, in Section~\ref{sec:rings}. 
Our new results on fluctuations of Kelvin waves and mutual friction in the spectrally-truncated Gross-Pitaevskii model are presented in Section~\ref{sec:new}. 
We recall the simple zero-temperature LIA model in Section~\ref{sec:newKVLIA}.
We present direct simulations of the finite-temperature spectrally-truncated GP model in Section~\ref{subsec:eqTGPKW}.
Noise and mutual friction terms are added to the simple zero-temperature LIA model in order to generate KW fluctuation in a Langevin effective model in  Section~\ref{sec:newNOISE}.
We relate the observed truncated GP fluctuation and relaxation to the simple Langevin effective model in Section~\ref{subsec:FiniteTemp_KW_in_GP}.
Finally, our results are discussed and some conclusions are given in
Sec.~\ref{sec:Conclusion}.

\section{Finite temperature states in truncated Gross-Pitaevskii equation}\label{sec:GP}

\subsection{The basic GPE model}\label{sec:basic}
The GP equation \cite{Gross61,Pitaevskii61} is the partial differential equation for the system's (complex) wave-function $\psi$ that reads
\begin{equation} 
    i\hbar\dertt{\psi}  =- \alps \gd \psi + \bet\lvert \psi\rvert ^2\psi .
    \label{Eq:GPE}
\end{equation}
It describes the dynamics of a zero-temperature dilute
superfluid BEC, $m$ is
the mass of the bosons, $g=4 \pi  \tilde{a} \hbar^2/m$,
($\tilde{a}$ is the $s$-wave scattering length). 
Equation  \eqref{Eq:GPE} admits three conservation laws corresponding to
the total (extensive) number of particles $N_p=\int_V n({\bf x}) \,d^3x$, the total energy $E=\int_V e({\bf x}) \,d^3x$ and
the momentum ${\bf P}=\int_V {\cal P}({\bf x}) \,d^3x$, with respective (intensive) densities:
\begin{eqnarray}
n&=& \lvert \psi\rvert ^2  ,\label{Eq:defN}\\
  e&= & \alps \lvert \nabla \psi \rvert ^2
    +\frac{g}{2}\lvert \psi\rvert ^4  \label{Eq:defH}\\
  {\bf \cal P}&=& \frac{i\hbar}{2}\left(
    \psi {\bf \nabla}\psib - \psib {\bf
    \nabla}\psi\right)\label{Eq:defP},
\end{eqnarray}
where the over-line denotes the complex conjugate.

When no zeros of $\psi$ are present, Eq.~(\ref{Eq:GPE}) can be easily mapped into hydrodynamic
equations of motion for a compressible irrotational fluid using the
Madelung transformation given by 
\begin{equation}
    \psi({\bf x},t)=\sqrt{\frac{\rho({\bf x},t)}{m}}\exp{[i
    \frac{m}{\hbar}\phi({\bf x},t)]},\label{Eq:defMadelung}
\end{equation}
where $\rho({\bf x},t)$ is the fluid density, and $\phi({\bf x},t)$ is
the velocity potential such that the fluid velocity is 
${\bf v}={\bf \nabla} \phi$. The Madelung transformation is singular on
the zeros of $\psi$, which can correspond to topological defects for the phase of the wave-function. 

In absence of vortices, Eq.~\eqref{Eq:GPE} can be linearised around a constant state
$\psi=A_0$. One obtains the Bogoliubov dispersion relation
\begin{equation} 
    \omega_B(k)=\sqrt{\frac{g k^2 \lvert A_0\rvert^2}{m}+\frac{\hbar^2 k^4}{4
        m^2}}.
    \label{eq:Bog}
\end{equation} 
The sound velocity is thus given by $c=\sqrt{g\lvert A_{0}\rvert^2/m}$, with
dispersive effects taking place for length scales smaller than the 
coherence length defined by
\begin{equation}
    \xi=\sqrt{\hbar^2/(2gm\lvert A_{\bf 0}\rvert^2) } ; \label{Eq:defxi}
\end{equation}
$\xi$ is also proportional to the radius of the vortex cores
\cite{Nore97a,Nore97b}.
When no vortex are present, the GP dynamics is similar to the compressible Euler dynamics, but with the addition of small-scale dispersive effects.
These dispersive effects are enough to not allow singularities that are present in the Euler case \cite{Nore1993}. In general, the 3D GPE is known to have regular solutions \cite{Gerard2006}.

When quantum vortices are present,  the velocity field can be directly obtained as
\begin{align}
    {\bf v} = \frac{\cal P} {n},
    \label{eq:vel}
\end{align}

The flow matches the behaviour of a classical, ideal,
and compressible potential fluid, except at
the topological singularities: the so-called quantum vortex lines with quantised Onsager-Feynman velocity circulation 
given by $\Gamma = \oint_C {\bf v}(\ell) \,d\ell = \frac{h}{m}$, where ${\bf v}$ is the superfluid velocity. 
The vorticity $\bf \omega=\nabla \times {\bf v}$ of the flow is thus given by
\begin{equation}
    {\bf \omega} ({\bf r}) = \Gamma \int d \zeta \frac{d {\bf s}'}{d \zeta}
    \delta^{(3)}({\bf r} - {\bf s}'(\zeta)),
    \label{vortdelta}
\end{equation}
where ${\bf s}(\zeta)$ is the position of the vortex lines and $\zeta$ the arc-length.
Note that the vorticity is non-zero although ${\bf v}={\bf \nabla} \phi$, because the phase is not singled value due to the topological defects produced by the zeros of $\psi$.

The simplest hydrodynamic excitations of a quantum vortex are dispersive helicoidal perturbations of the filament, known as Kelvin waves \cite{Pitaevskii61,Donnelly}. They were first studied by Lord Kelvin in 1880 \cite{Thomson_VibrationsColumnarVortex_1880} using the incompressible Euler equations. In the case of a hollow vortex core, such waves propagates with the dispersion relation
\begin{align}
    \label{kelvin}
    \omega_k^{\rm KW} = \frac{\Gamma}{2\pi a_0^2} \left(1 \pm
        \sqrt{1 + k a_0 \frac{K_0(ka)}{K_1(ka_0)}}\right) \stackrel{a_0k\ll1}{\approx}-\frac{\Gamma}{4\pi}k^2\left(\ln\frac{2}{a_0k}-\gamma_E\right),
\end{align}
where {$k$ is the wavenumber,} $a_0$ is a constant of the order of the vortex core radius, $\gamma_E\approx0.5772$ is the Euler-Mascheroni constant, and $K_0$ and $K_1$ are the modified Bessel functions. Note that their frequencies have the opposite sign with respect to the circulation $\Gamma$.

In the case of the GP model, the dispersion relation was first derived in the limit $a_0k\ll 1$ by Pitaevskii \cite{Pitaevskii61} and then improved by P.H. Roberts \cite{Roberts_VortexWavesCompressible_2003}. Robert's work allows to fix the constant of the core value to $a_0=1.1265\xi$. However, P.H.~Roberts also showed that in the opposite limit of large wave vectors, the vortex wave dispersion relation scales as $k^2$, which is not in agreement with the scaling $\omega_k^{\rm KW}\sim k^{1/2}$ for large $k$. In order to conciliate both scalings, reference \cite{Giuriato2020How} proposed a fit that matches both scaling laws. Such a fit was useful to perform analytical calculations and to analyse data. The whole dispersion relation of the GP is well represented by
\begin{align}
    \label{eq:omegaKWGP}
    \omega_k^{\rm GP} = \omega_k^{\rm KW}\times\left[1 + \epsilon_{\frac 12} (a_0 k)^{1/2}+\epsilon_{ 1} (a_0 k)+\frac 12 (a_0 k)^{3/2}   \right],
\end{align}
where $\epsilon_{\frac 12}=-0.20$ and $\epsilon_{1}=0.64$ are two fitting parameters. The last term of the expansion is fixed to match the $a_0k\gg1$ GP asymptotic limit. Finally, note that in the simplest KW description given by the Local-Induction-Approximation (LIA) \cite{DaRios_SulMotoLiquido_1906} (see also Section \ref{sec:newKVLIA}), the $log$-term is ignored and treated as a constant. 

\begin{figure}[h]%
    \centering
    \includegraphics[width=.8\textwidth]{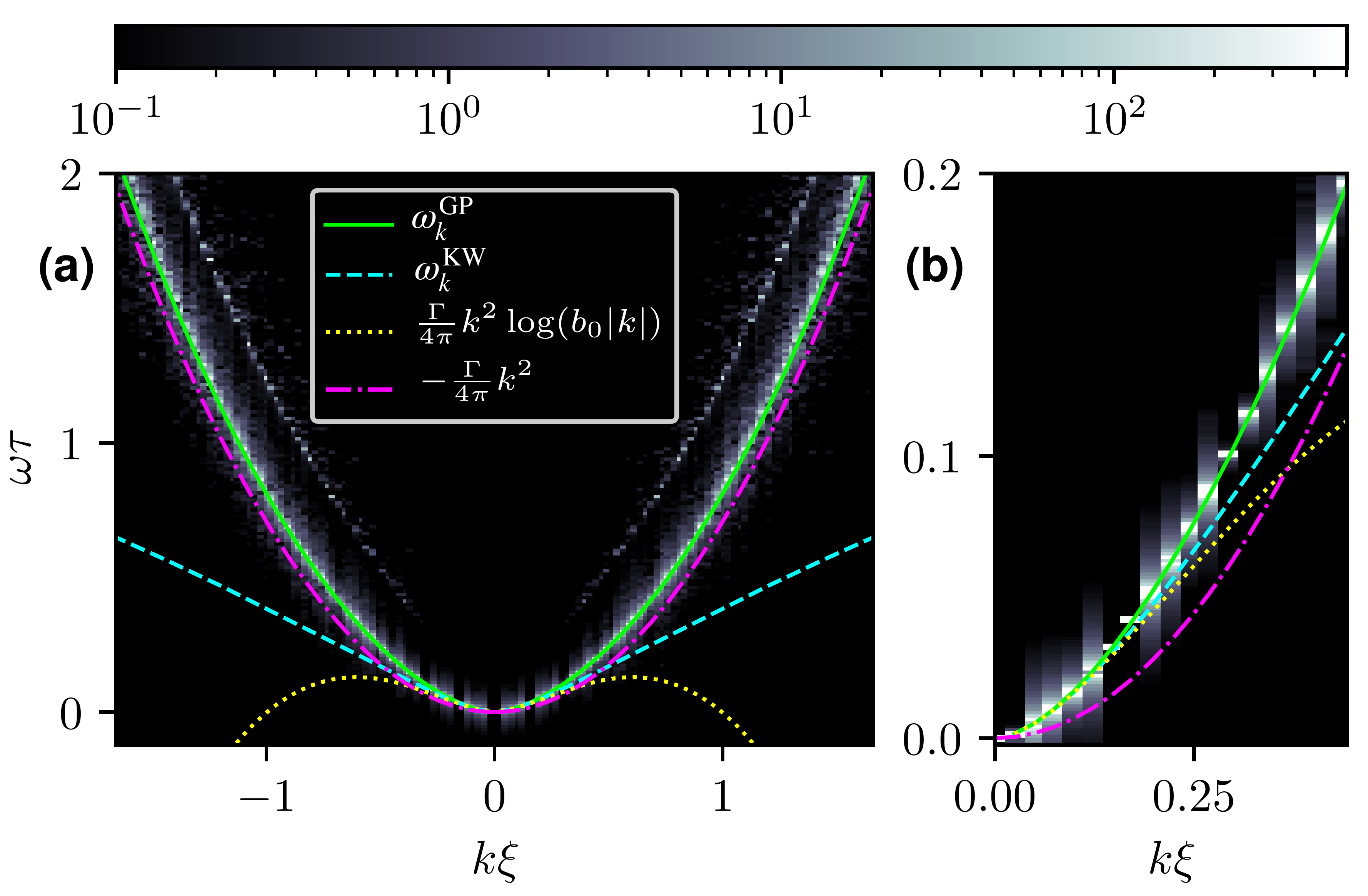}
    \caption{Figure reproduced from reference \cite{Giuriato2020How}. \textbf{(a)} Kelvin wave dispersion relation obtained from GP simulations. All theoretical predictions are also displayed for comparison. \textbf{(b)} A zoom for $k\xi\ll1$ where the logarithmic correction can be appreciated. {Details on the numerical simulations can be found in \cite{Giuriato2020How}.}}\label{fig:KWdispRelation}
\end{figure}

Figure \ref{fig:KWdispRelation}, reproduced from reference \cite{Giuriato2020How}, shows the measured KW dispersion relation in GP simulations and different theoretical predictions.

\subsection{Truncation and conserved quantities}\label{sec:trunc}
The truncated Gross-Pitaevskii (TGP) equation is obtained from the GP model by truncating the Fourier transform of the wave-function $\psi$: $\hat{\psi}_{\bf k}\equiv0$ for $\lvert{\bf k}\rvert>\kmax$ \cite{Davis01,Proukakis08}.
Introducing the Galerkin projector $\mathcal{P}_{\rm G}$ that reads in Fourier space $\mathcal{P}_{\rm G} [ \hat{\psi}_{\bf k}]=\theta(\kmax-\lvert{\bf k}\rvert)\hat{\psi}_{\bf k}$ with $\theta(\cdot)$ the Heaviside function, the TGP equation  explicitly reads
\begin{equation}
i\hbar\dertt{\psi}  =\mathcal{P}_{\rm G} [- \alps \gd \psi + \bet \mathcal{P}_{\rm G} [\lvert\psi\rvert^2]\psi ].
\label{Eq:TGPEphys}
\end{equation}

{The truncation of the GP equation preserves its Hamilonian structure. The correponding Hamiltonian is $H_{\rm TGP}=\int \mathcal{P}_{\rm G}\left[ \alps \lvert \nabla \psi\lvert^2 + \frac\bet2\left(\mathcal{P}_{\rm G} [\lvert\psi\rvert^2] \right)^2\right] d^3x$. The outest projector can be omitted if we assume that initially $\mathcal{P}_{\rm G}\psi=\psi$. Equation \eqref{Eq:TGPEphys} can be thus rewritten in the cannonical way as  $i\hbar\dertt{\psi}=\frac{\delta H_{\rm TGP}}{\delta \psi^*}$. It follows that energy and mass are conserved by \eqref{Eq:TGPEphys} as the time and $U(1)$ invariances are preserved in $H_{\rm TGP}$ Conservation of momentum is also preserved when} standard Fourier pseudo-spectral methods are used, provided that they are dealiased using the $2/3$-rule ($\kmax=2/3\times M/2$ \cite{Got-Ors} at resolution $M$).
In this scheme, to achieve conservation mass and momentum are respectively evaluated as in  \eqref{Eq:defN}, \eqref{Eq:defP} and the energy \eqref{Eq:defH} must be evaluated as in $e=\alps \lvert \nabla \psi \rvert^2 +\frac{g}{2}[\mathcal{P}_{\rm G}\lvert\psi\rvert^2]^2$.
Note that global momentum conservation is mandatory to correctly describe vortex-normal fluid interactions. When the nonlinear term in Eq.\eqref{Eq:TGPEphys} is written, as in  \cite{Davis01}, $\mathcal{P}_{\rm G} [\lvert\psi\rvert^2 \psi]$ momentum conservation requires that the dealiasing must be performed at $\kmax=M/4$ (see reference \cite{Krstulovic11} and \cite{KrstulovicHDR} for details).

{Note that when the wavefunction $\psi$ is an analytical function (or more regular), it decreases very rapidly in Fourier space with the magnitude of the wave number (typically exponentially or faster). In that case, and provided that $\kmax$ is much larger than the largest active wavenumber of $\psi$, the GP and TGP equations coincide. This criterium is called spectral accuracy and should be fulfilled while solving a partial differential equation. On the contrary, the TGP model allows for solutions which do not decay at large wave numbers, which typically correspond to the thermal states discussed in the next section. As fields do not decrease for large wave numbers, they are not differentiable at any point. Formally speaking, the TGP is not a partial differential equation but a high-dimensional ordinary differential equation.}

\subsection{Equivalence micro and grand-canonical thermalisation}\label{sec:equiv}

Microcanonical equilibrium states are produced by integration of TGP equation for very long times \cite{Davis01,Connaughton_2005,Berloff07}. 
On the other hand, grand canonical states are defined by the probability distribution 
\begin{equation}
\mathbb{P}_{\rm st}[\psi]=\mathcal{Z}^{-1}\exp[{-\beta E-\mu N)}] \label{eq:GC}
\end{equation}
directly expressed in terms of the temperature $\beta^{-1}$ and the chemical potential $\mu$ (instead of the energy $E$ and the number of particles $N$ in a microcanonical framework). These states can be efficiently obtained  by constructing a stochastic process that converges to a realisation with the probability $\mathbb{P}_{\rm st}[\psi]$ \cite{Krstulovic11}. This process is defined by a Langevin equation consisting in a stochastic Ginbzurg-Landau equation (SGLE):
\begin{equation}
\hbar\dertt{\psi} =\mathcal{P}_{\rm G}  \left[\alps \gd \psi +\mub \psi - \bet\mathcal{P}_{\rm G} [\lvert\psi\rvert^2]\psi\right]  +\sqrt{\frac{2 \hbar}{V\beta }}  \mathcal{P}_{\rm G} \left[\zeta({\bf x},t)\right] 
\label{Eq:SGLRphys},\hspace{3mm}
\end{equation}
where the white noise  $\zeta({\bf x},t)$ satisfies $\langle\zeta({\bf x},t)\zeta^*({\bf x'},t')\rangle=\delta(t-t') \delta({\bf x}-{\bf x'})$, $\beta$ is the inverse temperature and $\mu$ the chemical potential. 
Using this algorithm in \cite{Krstulovic11} the microcanonical and grand canonical ensembles were shown to be equivalent and the condensation transition reported in \cite{Davis01,Connaughton_2005} identified with the standard second order $\lambda$-transition. 

\begin{figure}[htbp]
\centering
\includegraphics[height=5cm]{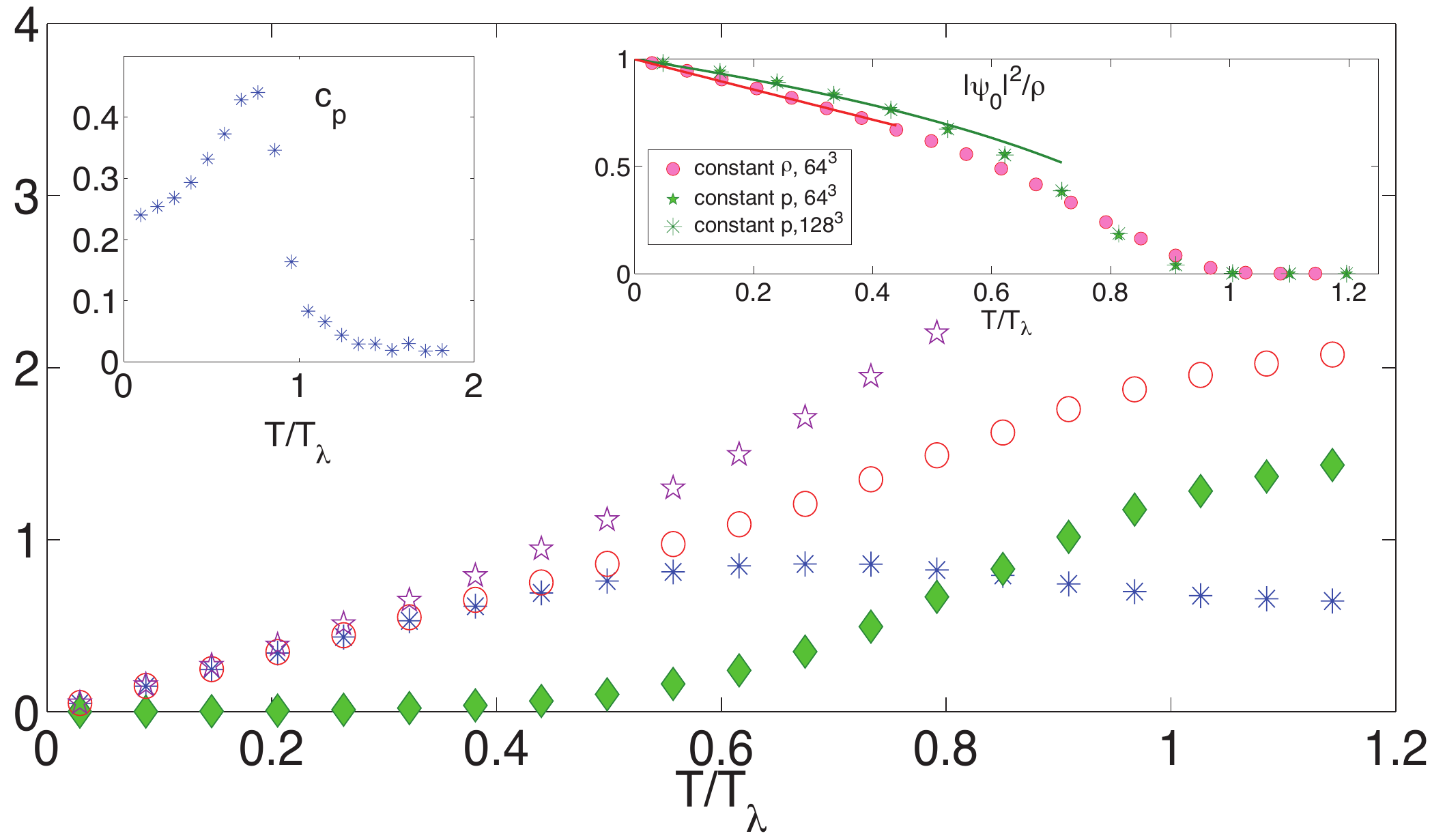}
\caption{This figure, reproduced from \cite{Krstulovic11a}, shows the temperature dependence of various energies: $E_{\rm kin}^{\rm c}$ (stars), $E_{\rm kin}^{\rm i}$ (diamonds), $E_{\rm kin}$ (circles), and $E_{\rm q}+E_{\rm int}$ (pentagrams) at constant density. The insets show the temperature dependence of the condensate fraction $\lvert\psi_{\bf 0}\rvert^2/\rho$ (right) and the specific heat $c_{p}=\left.\frac{\partial H}{\partial T}\right\rvert_{p}$ at a resolution of $128^3$ (left). Note that $E_{\rm kin}^{\rm i}$ denotes the presence of vortices (see text). {Details on the numerical simulations can be found in \cite{Krstulovic11a}.}}
\label{Fig:Scan}
\end{figure}

When numerically simulating Eq.\eqref{Eq:SGLRphys}, $\mu$ is adjusted to fix the density. The inverse temperature is normalized as $\beta=\mathcal{N}/VT$, where $\mathcal{N}$ is the total number of Fourier modes and $V$ the system volume. With this choice of parametrization the $\lambda$-transition temperature $T_\lambda$ is independent of $\mathcal{N}$. Data from SGLE and low-temperature calculations obtained in reference \cite{Krstulovic11} are shown in figure~\ref{Fig:Scan}.

The temperature dependence of the different energies is also displayed in Fig.\ref{Fig:Scan}.
Note that the first term in equation \eqref{Eq:defH} that defines the energy can be decomposed (as done in \cite{Nore97a,Nore97b}) into incompressible $E_{kin}^{\rm i}$, and compressible $E_{kin}^{\rm c}$ kinetic energies and into quantum energie $E_{q}$. The last term  in equation \eqref{Eq:defH} corresponds  to the internal energy $E_{int}$. Observe that $E_{\rm kin}^{\rm i}$ vanishes at temperatures $T/T_{\lambda}\lesssim1/2$. 
At low temperature equipartition of energy between $E_{\rm kin}$ and $E_{\rm q}+E_{\rm int}$ is apparent. Such a behaviour suggest that thermally excited quantum vortices are important for temperatures larger than $T_\lambda/2$.

Adding $-{\bf v_{\rm n}}\cdot {\bf P}$ as a Lagrange multiplier to the grand canonical distribution \eqref{eq:GC} and correspondingly a term
$i\hbar\,{\bf v_{\rm n}}\cdot{\bf\nabla} \psi$ in \eqref{Eq:SGLRphys} induces an asymmetry in the repartition of thermal waves and generates non-zero momentum states. These states do not generally correspond to a condensate moving at velocity ${\bf v_{\rm s}}={\bf v_{\rm n}}$ because ${\bf v_{\rm s}}$ is the gradient of a phase and takes discrete values for finite size systems.
Equilibrium states states with nonzero values of the counterflow ${\bf w}={\bf v_{\rm n}-v_{\rm s}}$ are generated in this way.

All the SGLE equilibrium used in this work have a condensate at rest (${\bf v_{\rm s}=0}$) and therefore ${\bf v_{\rm n}}={\bf w}$.

\section{Vortex lines at finite temperature}\label{sec:mutual}
In this section, we review the results on mutual friction obtained in references \cite{Krstulovic11} and reference \cite{Krstulovic11b}.

\subsection{Mutual friction in the truncated Gross-Pitaevskii model} \label{sec:lines}

We start by recalling the standard Hall-Vinen phenomenological model for the vortex line velocity ${\bf v}_{\rm L}$ \cite{Donnelly,HallVinen1_1956,HallVinen2_1956} that reads:
\begin{equation} 
{\bf v}_{\rm L}={\bf v}_{\rm sl}+\alpha {\bf s'}\times({\bf v}_{\rm n}-{\bf v}_{\rm sl})-\alpha'{\bf s'}\times[{\bf s}'\times({\bf v}_{\rm n}-{\bf v}_{\rm sl})],\label{Eq:VortexDyn}
\end{equation}
where ${\bf s}'$ is the tangent of the vortex line, ${\bf v}_{\rm sl}={\bf v}_{\rm s}+{\bf u}_{\rm i}$ is the local superfluid velocity with  ${\bf u}_{\rm i}$  the self-induced vortex velocity and ${\bf v_{\rm n}}$ the normal velocity. The mutual friction coefficients in Eq.~\eqref{Eq:VortexDyn} are typically written as $\alpha=B\rho_{\rm n}/2\rho, \alpha'=B'\rho_{\rm n}/2\rho$ where $B$ and $B'$ are order-one and weakly temperature-dependent. 

In reference \cite{Krstulovic11b} we used the TGP model to study the effect of mutual friction on an array of alternate-sign straight vortices $\psi_{\rm lattice}$, which is an exact stationary solution of the GP equation (for details on this initial data, see \cite{Krstulovic11} and also \cite{Nore94}). We set an initial condition where the vortices were separated by a distance $L/2$ ($L$ being the system size) and thus they can be considered isolated for a large enough box. An equilibrium state $\psi_{\rm eq}$ was also generated with the SGLE \eqref{Eq:SGLRphys} with counterflow $v_{\rm n}$ perpendicular to the vortices. The initial condition $\psi=\psi_{\rm lattice}\times\psi_{\rm eq}$ was then evolved with the TGP equation.

\begin{figure}[htbp]
\begin{center}
\includegraphics[height=7cm]{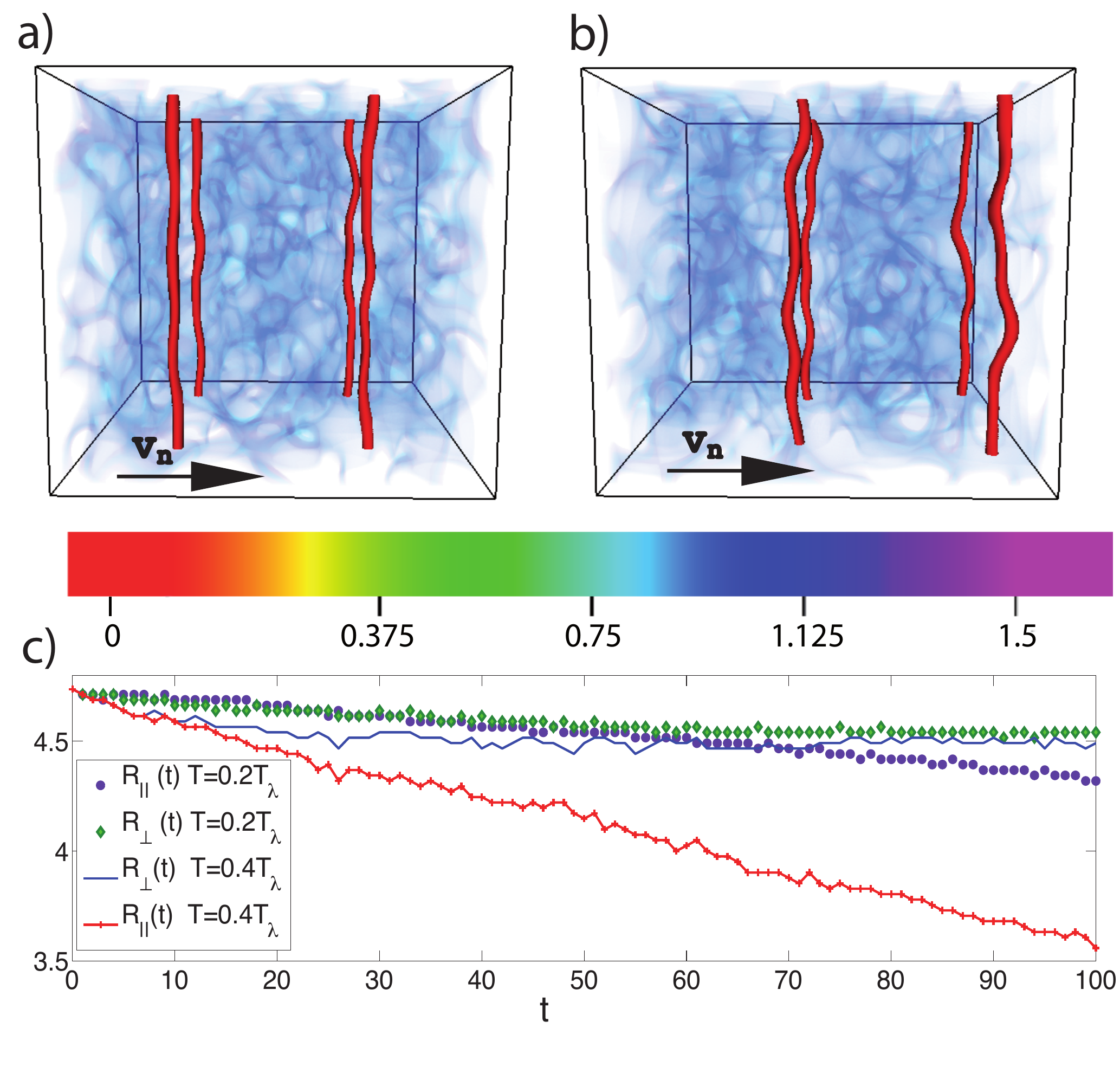}
\caption{The figure, reproduced from reference \cite{Krstulovic11b}, displays three panels: a and b show the density distribution at $t=0$ and $t=100$, respectively, for the lattice configuration (in red) with $T=0.4,T_\lambda$ and $v_{\rm n}=0.4$, where the blue clouds correspond to density fluctuations. Panel c shows the {averaged} positions $(R_{\parallel},R_{\perp})$ of a single vortex for different temperatures ($T=0.2,T_\lambda$ and $T=0.4,T_\lambda$) and $v_{\rm n}=0.4$. The resolution used in this figure was $64^3$. {Details on the numerical simulations and method used can be found in \cite{Krstulovic11b}.} }
\label{Fig:Lattice}
\end{center}
\end{figure}

Figure \ref{Fig:Lattice}.a displays $3$D visualisations of the density at $t=0$ and $t=100$, where the displacement of the lattice is apparent (figure reproduced from \cite{Krstulovic11b}). 
The temporal evolution of the (parallel and perpendicular to ${\bf v}_{\rm n}$) position of a vortex $(R_{\parallel},R_{\perp})$  
are presented on Fig.\ref{Fig:Lattice}.c for $T=0.2\,T_\lambda$, $T= 0.4\,T_\lambda$ and $v_{\rm n}=0.4$. 
The counterflow-induced vortex velocity clearly depends on the temperature. 
$R_{\parallel}$ has a linear behaviour, that allows to directly measure the parallel velocity $v_{\parallel}$. The temperature dependence of $v_{\parallel}/v_{\rm n}$ is presented later on Fig.~\ref{Fig:alphap} for different values of $v_{\rm n}$ and $\xi$. This behaviour is consistent with the standard phenomenological model for the vortex line velocity ${\bf v}_{\rm L}$ \eqref{Eq:VortexDyn}, as Eq.~\eqref{Eq:VortexDyn} applied to a straight vortex with $v_{\rm n}$ perpendicular to the vortex and $v_{\rm s}=0$ yields $\alpha'=v_{\parallel}/v_{\rm n}$. 

\subsection{Slowdown effect on vortex rings induced by thermal fluctuations}\label{sec:rings}
Let us now rapidly review the interaction of vortex rings and counterflow. 
The Biot-Savart self-induced velocity of a perfectly circular vortex ring of radius $R$ is given by
\begin{equation}
u_{\rm i}=\frac{\hbar}{2m}\frac{C(R/\xi)}{R}\,,\hspace{5mm} C(z)=\ln{(8z)}-a \label{Eq:ui}
\end{equation} 
where $a$ is a core model-depending constant \cite{Donnelly}. 

\begin{figure}[htbp]
\begin{center}
\includegraphics[height=6cm]{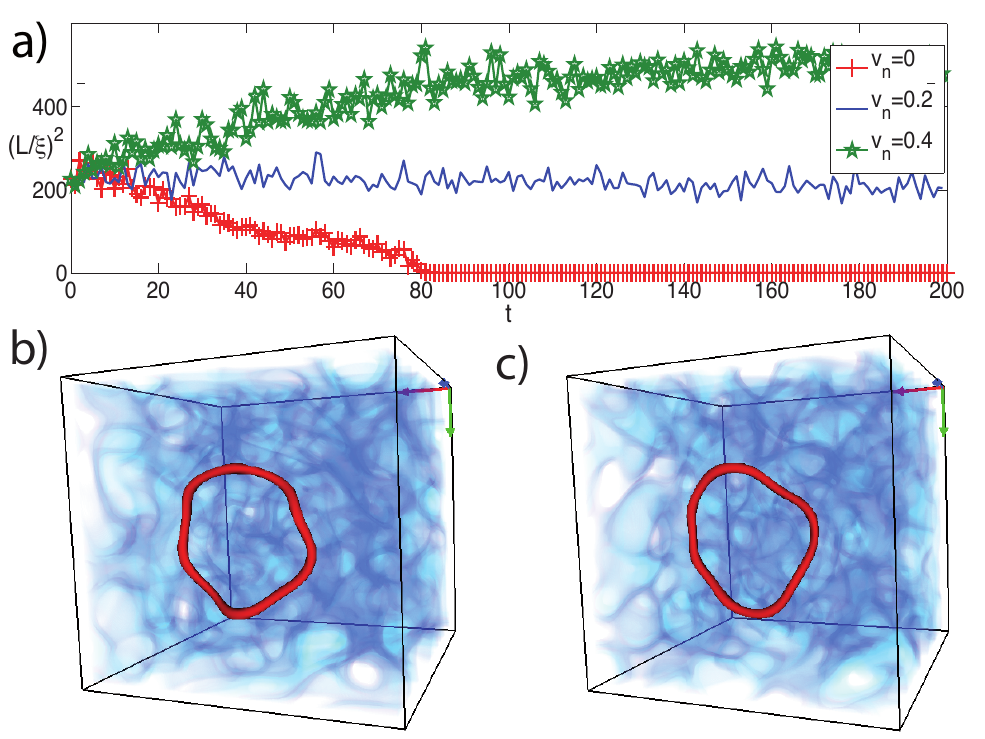}
\caption{Figure reproduced from reference \cite{Krstulovic11b}. a) The (squared) length of a vortex ring at different values of counterflow $v_{\rm n}$ as a function of time (temperature $T=0.4,T_{\lambda}$ and initial radius $R=15\xi$). b-c) $3$D visualisations of a vortex ring ($R=20\xi$) and density fluctuations at $t=18$, $19$ with $T=0.4,T_{\lambda}$ and resolution $64^3$. The same colour bar as in Fig.\ref{Fig:Lattice} is used. Thermally-excited Kelvin waves can be seen. {Details on the numerical simulations can be found in \cite{Krstulovic11b}.}}
\label{Fig:3}
\end{center}
\end{figure}

Equation \eqref{Eq:VortexDyn} with $v_{\rm n}$ perpendicular to the ring, and $v_{\rm s}=0$ yields the radial velocity $\dot{R}=-\alpha(u_{\rm i}-v_{\rm n})$. 
Using the TGP model, Berloff and Youd \cite{Berloff07} studied the finite temperature evolution of a ring without counterflow ($v_{\rm n}=0$) and observed a contraction of vortex rings compatible with \eqref{Eq:VortexDyn}. To study the influence of counterflow, in reference \cite{Krstulovic11b} we prepared an initial condition $\psi=\psi_{\rm ring}\times\psi_{\rm eq}$ in the same way as above for the vortex lattice. The temporal evolution of the (squared) vortex length of a ring of initial radius $R=15\xi$ at temperature $T=0.4\,T_{\lambda}$ and $v_{\rm n}=0$, $0.2$ and $0.4$ is reproduced on Fig.\ref{Fig:3}.a. 
The Berloff-Youd contraction \cite{Berloff07} is apparent in absence of counterflow (bottom curve). The temperature dependence of the contraction, related to the $\alpha$ coefficient in Eq.\eqref{Eq:VortexDyn}, also quantitatively agrees with their published results (data not shown).

When the counterflow was large enough, a dilatation of vortex rings was was instead observed (top curve on Fig.\ref{Fig:3}.a).
Such a dilatation (a hallmark of counterflow effects) is expected \cite{Donnelly} to correspond to a change of sign of ${\bf v}_{\rm n}-{\bf v}_{\rm sl}$ in Eq. \eqref{Eq:VortexDyn}. 
The predictions of Eq.\eqref{Eq:VortexDyn} unexpectedly turned out to be quantitatively wrong. Indeed, using Eq.\eqref{Eq:ui} in the conditions of Fig.\ref{Fig:3}.a one finds that ${\bf v}_{\rm sl}={\bf u}_{\rm i}=0.39$
which is significantly larger than normal velocity $v_{\rm n}=0.2$ around which
dilatation starts to take place (see the middle curve in Fig.\ref{Fig:3}.a).
Equation \eqref{Eq:VortexDyn} prediction for the longitudinal velocity $v_{L}=(1-\alpha')u_{\rm i}+\alpha'v_{\rm n}$ was also unexpectedly found wrong. 
Using the value of $\alpha'$ determined above on the vortex array, one finds $v_{L}\sim 0.98 u_{\rm i}$ and from Eq.\eqref{Eq:ui} one finds for $v_{L}$ the value $0.38$ that is larger than the measured value $v_L=0.23$.

\begin{figure}[htbp]
\begin{center}
\includegraphics[height=3.5cm]{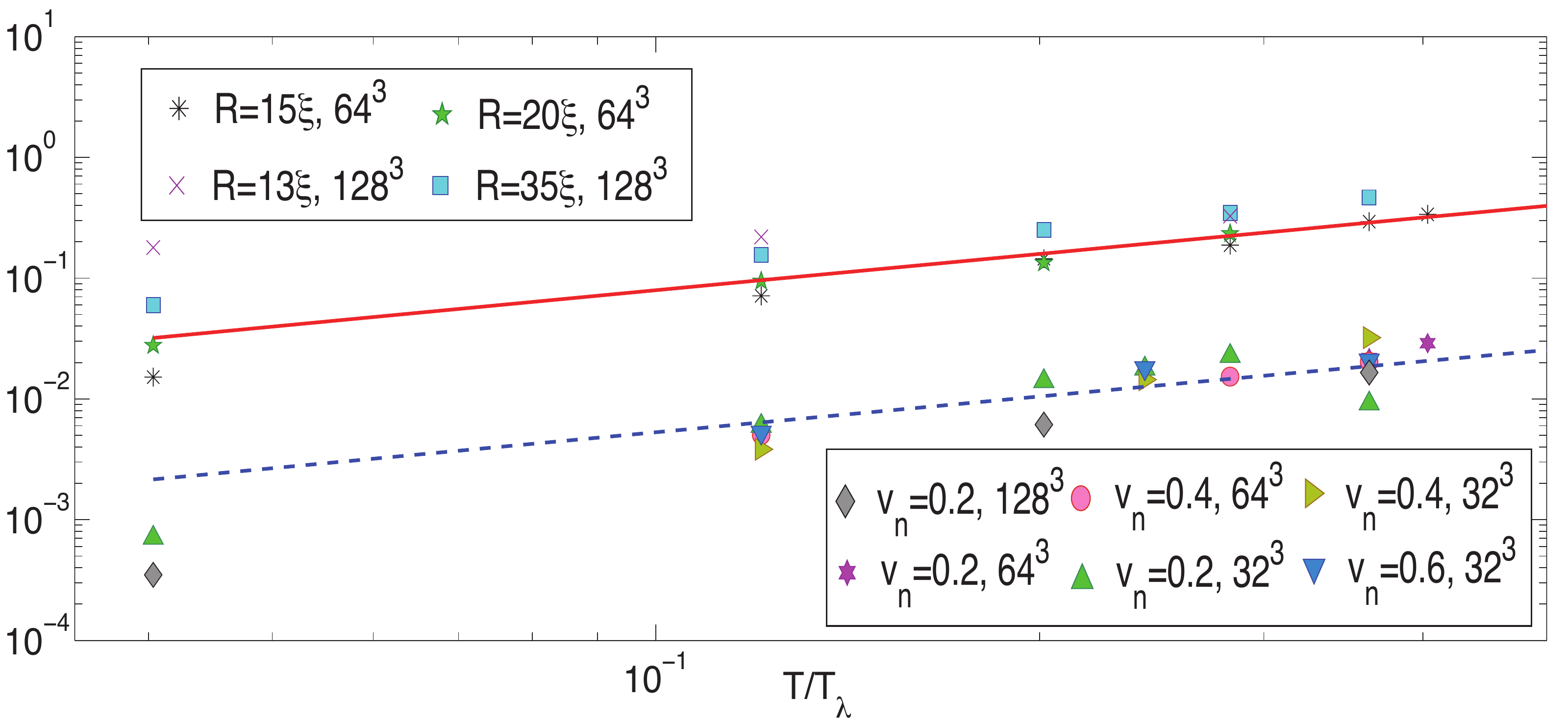}
\caption{This figure is reproduced from reference \cite{Krstulovic11b}. It shows the temperature dependence of the counterflow-induced lattice velocity $v_{\parallel}/v_{\rm n}$ (bottom) and ring slowdown $\Delta v_{L}/u_{\rm i}$ (top) obtained with $v_{\rm n}=0$. The dashed line represents the prediction of Eq.\eqref{Eq:VortexDyn} with $\alpha'=0.83\rho_{\rm n}/2\rho$, while the solid line shows the prediction of anomalous slowdown by Eq.\eqref{Eq:AnEffect} with $R=20\xi$ at various resolutions. {Details on the numerical simulations can be found in \cite{Krstulovic11b}.} }
\label{Fig:alphap}
\end{center}
\end{figure}

This anomaly of the ring velocity $v_L$ is also present in the absence of counterflow ($v_{\rm n}=0$) where Eq.\eqref{Eq:VortexDyn} predicts that $\alpha'$ should be given by $\Delta v_{L}/u_{\rm i}\equiv(u_{\rm i}-v_{L})/u_{\rm i}$. 
The temperature dependence of $\Delta v_{L}/u_{\rm i}$, reproduced from \cite{Krstulovic11b}, is displayed on Fig.~\ref{Fig:alphap} (top curve). 
Observe that $\Delta v_{L}/u_{\rm i}$ is one order of magnitude above the transverse mutual friction coefficient $\alpha'$ measured on the lattice. 

This unexpected behaviour was related to the presence of thermally excited Kelvin waves in reference \cite{Krstulovic11b}. It was known that KWs present in vortex rings slowdown the ring propagation velocity by a purely hydrodynamical effect \cite{Kiknadze:2002p3905,Barenghi:2006p3901}. As KWs are naturally excited at finite temperatures, they thus could be used to explain this unexpected observed effect. We now reproduce the calculations of \cite{Krstulovic11b} of the thermally-induced anomaly to the velocity $v_{\rm a}$ induced on a vortex ring by a single a Kelvin wave of (small) amplitude $A$ and (large) wavenumber $N_{\rm K}/2\pi R$ obtained in the LIA \cite{Kiknadze:2002p3905} and Biot-Savart \cite{Barenghi:2006p3901} frameworks.
The translation velocity $v_{\rm a}$ obtained in the framework of LIA reads (see Eq. (26) of \cite{Kiknadze:2002p3905})
\begin{equation}
v_{\rm a}= u_{\rm i}(1-A^2N_{\rm K}^2/R^2+3A^2/4R^2)\label{Eq:AneffKik}
\end{equation}
where  $u_{\rm i}$ is the (undisturbed) ring velocity \eqref{Eq:ui}.

The TGP model naturally includes thermal fluctuations that excite Kelvin waves, as apparent on Fig.\ref{Fig:3}.b-c. 
We assume that the slowing down effect of each individual Kelvin wave is additive and that the waves populate all the possible modes. 
Kelvin waves being bending oscillations of the the quantised vortex lines their wavenumber must satisfy $k \le k_\xi= 2 \pi / \xi$. The total number of thermally excited Kelvin waves is thus
$\mathcal{N}_{\rm Kelvin}\approx R\,k_\xi$.

The amplitude term $A^2N_{\rm K}^2/R^2$ in \eqref{Eq:AneffKik} can be obtained by simple equipartition arguments. The energy of a (perfect) ring is $E=\frac{2 \pi^2\rho_{\rm s}\hbar^2}{m^2}R[C(R/\xi)-1 ]$, with $\rho_{\rm s}$ the superfluid density \cite{Donnelly}.  A Kelvin wave produces a variation of the ring length $\Delta L=\pi A^2 N_{\rm K}^2/R$. Its energy can thus be estimated as
$\Delta E=\frac{dE}{dR}\frac{\Delta L}{2\pi}.$
Assuming $\Delta E=\beta^{-1}$ yields, at low temperature where $\rho_{\rm s}\approx\rho$, $A^2N_{\rm K}^2/R^2=m^2\beta^{-1}/\pi^2\rho\hbar^2R\,C(R/\xi)$. Note that this formula predicts a $UV$-convergent r.m.s amplitude that is in good agreement with TGP data, with values small enough to avoid self-reconnections of the ring.
Replacing $A^2/R^2$ in Eq.\eqref{Eq:AneffKik}, the dominant effect is obtained by summing up to $\mathcal{N}_{\rm Kelvin}$ and it finally reads:
\begin{equation}
\frac{\Delta v_{L}}{u_{\rm i}}\equiv\frac{u_{\rm i}-v_{\rm a}}{u_{\rm i}}\approx\frac{\beta^{-1}m^{2}}{\pi^2 \rho \hbar^{2}C(R/\xi)}k_\xi\label{Eq:AnEffect}.
\end{equation}
The thermally-induced anomalous slowdown \eqref{Eq:AnEffect} is in good agreement with the TGP data displayed on Fig.\ref{Fig:alphap}.

\section{Fluctuation and relaxation of Kelvin waves in thermal equilibrium}\label{sec:new}
When the standard LIA description of Kelvin waves is used together with the Hall-Vinen  \cite{HallVinen1_1956,HallVinen2_1956} phenomenological description of mutual friction, the normal fluid velocity appears naturally as an external variable that is given independently of the vortex positions.
We have seen that the TGP model naturally includes thermal fluctuations that excite Kelvin waves.  Such random thermal perturbations are physically expected to occur in nature and not only in the TGP model. 

In this section, we review the description of KWs in the LIA model, their expected energy equipartion at finite temperatures, and build a simple Langevin model of the effect of thermal perturbations by including random fluctuations in the normal fluid velocity.
We will be specially interested in the fluctuations and relaxation of thermally excited Kelvin waves. 
 
\subsection{Description of Kelvin waves at $T=0$ using the Local-Induction-Approximation} \label{sec:newKVLIA}

The simplest description of Kelvin wave dynamics is given by the local induction approximation (LIA)~\cite{DaRios_SulMotoLiquido_1906}, in which only local contributions of the Biot-Savart integrals in the Schwarz vortex filament model \cite{Schwarz_85,Schwarz_88} are considered. In the zero-temperature limit, the LIA equations are simply given by
\begin{equation}
    \dot{{\bf s}}=-\frac{\Gamma \Lambda}{2\pi}{\bf s}'\times {\bf s}''
\end{equation}
where ${\bf s}$ denotes, as in \eqref{Eq:VortexDyn}, the parametrisation of a vortex line. The vortex circulation is given by $\Gamma$ and the constant $\Lambda=\log{(\ell/a_0)}$, with $\ell$ the inter-vortex distance and $a_0$ the vortex core size.

In the particular case of small amplitude Kelvin waves, the vortex parametrisation takes a simpler form using Cartesian coordinates. For a vortex line aligned along the $z$-direction, the vortex can be parametrised as ${\bf s}(z)=(X(z),Y(z),z)$. Furthermore, by introducing the complex amplitude $s(z)=X(z)+iY(z)$, and taking the limit of small amplitudes, the LIA equations become 
\begin{equation}
    i\dot{s}=-\frac{\Gamma \Lambda}{4\pi}\frac{\partial^2s}{\partial z^2}.
\end{equation}
Note that the previous equation is equivalent to the (linear) Schrödinger equation. In particular, note that in the LIA approximation, KWs have a dispersion relation 
\begin{equation}
    \omega_k^{\rm LIA}=-\frac{\Gamma \Lambda}{4\pi}k^2,
\end{equation}
where $k$ is the wave vector. Although the LIA equations fails to give the correct large-scale asymptotic dispersion relation \eqref{kelvin}, it is a useful theoretical tool to perform analytical calculations and comparison with the GP model \cite{Giuriato_InteractionActiveParticles_2019,Giuriato2020How}.

The LIA equations have a Hamiltonian structure in which energy is proportional to the total vortex length. In the case of small KW amplitudes, the energy of the filament is given by
 \begin{equation}
     H_{\rm LIA}=\frac{\Gamma^2 \Lambda}{4\pi} \int \left\lvert\frac{\partial s}{\partial z}\right\lvert^2 dz= -\Gamma\sum_k \omega_k^{\rm LIA} \lvert \hat{s}_k\lvert^2,
 \end{equation}
where $\hat{s}_k$ is the Fourier transform of $s(z)$, and for simplicity we have considered a periodic system in the $z$-direction.

At finite temperatures, and in absence of external driving, one naturally expects that KWs will be in equilibrium with the thermal bath. More specifically, in the classical approximation, vortex line excitations should obey the Gibbs distribution $\{s(z)\}\sim \exp{\{-H_{\rm LIA}/T\}}$, where $T$ is the temperature of the thermal bath. As $H_{\rm LIA}$ is quadratic in $s$, every Fourier mode will be Gaussian and will be in equipartion. It follows that the thermal KW spectrum is given by 
\begin{equation}\label{eq:thermalKWspectrum}
    n_k=\lvert \hat{s}_k\lvert^2+\lvert \hat{s}_{-k}\lvert^2\propto \frac{T}{-\Gamma\omega^{\rm LIA}_k}\sim\frac{T}{k^2}.
\end{equation}

The previous considerations are also valid using a more realistic description of linear KWs. For a different model, it is enough to replace ${\omega_k^{\rm LIA}}$ by a more accurate expression of the KW dispersion relation, for instance the one given by Eq.~\eqref{eq:omegaKWGP}.

\subsection{Finite-temperature Kelvin wave Gross-Pitaevskii simulations}\label{subsec:eqTGPKW}

In order to check prediction \eqref{eq:thermalKWspectrum}, we study the finite-temperature evolution of an initial non-equilibrium distribution of KWs using the TGP Eq.~\eqref{Eq:TGPEphys}. For a given temperature $T$, which value will be expressed in terms of the condensation transition temperature $T_\lambda$, we first prepare a thermal state $\psi^{\rm thermal}$. Then, we also prepare a wave function $\psi^{\rm KW}$ containing four (to ensure periodicity) almost straight vortices satisfying
\begin{equation}
s^{\rm ini}(z)=\sum_{k={-k_m}}^{k_m} Ae^{ikz +i \phi_k},\label{eq:vortexWithKW}
\end{equation}
where $\phi_k$ are random phases and $A$ is the KW amplitude. The initial condition is prepared in the same manner as in \cite{Krstulovic_KelvinwaveCascadeDissipation_2012}. Finally, the initial condition $\psi=\psi^{\rm thermal}\times\psi^{\rm KW}$ is evolved with the TGP Eq.~\eqref{Eq:TGPEphys}. During the evolution, we track the four vortex lines of the periodic box. {In general, tracking KW requires high precision in order to capture small fluctuations. In the case of GP (not truncated), one can use the regularity of the field and a Newton–Raphson scheme to find the nodal lines of the wave function \cite{Krstulovic_KelvinwaveCascadeDissipation_2012,Villois_VortexFilamentTracking_2016}. This approach is particularly accurate in pseudo-spectral codes thanks to the spectral accuracy of such solvers. As for finite temperature states, the fields are not differentiable. This method and others based on interpolation are not thus applicable. Instead, in this work, we use a simpler scheme where the vortex is determined by finding the minimum the density, after filtering the fields to reduce small-scale fluctuations. As we will say later, this simple method allows for determining statistical quantities with good precision.}


In the numerics we use the pseudo-spectral code FROST \cite{KrstulovicHDR} with a periodic box of size $L\times L\times 2L$ using $128\times128\times256$ collocation points and a Runge-Kutta-4 time-marching scheme. Thermal states are prepared at constant density $\rho=1$. At zero-temperature, the healing length is set to $\xi=1.5L/128$. Furthermore, as the vortex filaments are highly fluctuating quantities, for each temperature studied in this work, we perform 27 different realisations to improve statistics. As the four vortices can be considered to be statistically independent, this results in a statistical ensemble of more than $100$ vortex lines for each temperature. Finally, we prepare an out-of-equilibrium vortex configuration $\psi^{\rm KW}$, {with an almost straight vortex as in Eq.\eqref{eq:vortexWithKW}} with $k_m=20k_0$, with $k_0=\pi/L$. We filter the fields at wave number $k_f=15k_0$ to track the vortex lines. The amplitude of the KWs is set to $5\xi/\sqrt{2}$.

\begin{figure}[h]%
    \centering
    \includegraphics[width=0.9\textwidth]{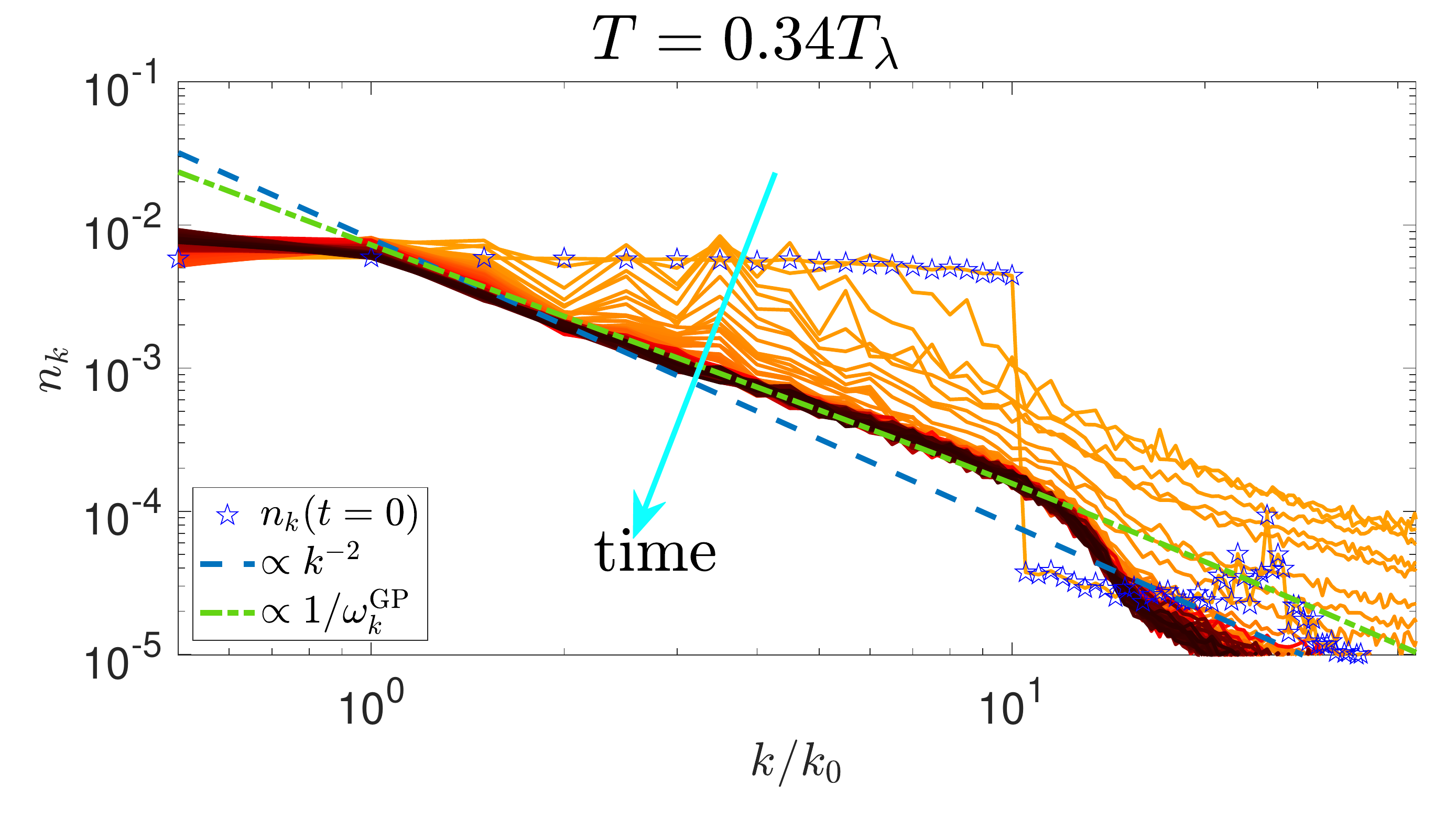}
    \caption{Temporal evolution of the Kelvin wave spectrum $n_k$. The initial condition is highlighted with pentagrams and the cyan and green dashed line show the LIA $k^{-2}$ scaling and $1/\omega_k^{\rm GP}$ prediction \eqref{eq:omegaKWGP}. $k_0$ is the smallest wavenumber of the simulation.}\label{fig:thermalspec}
    \end{figure}

Figure \ref{fig:thermalspec} displays the temporal evolution of the KW spectrum at a temperature $T=0.34T_\lambda$. 
Initially, the spectrum is flat and drops at $k=10k_0$ as set by the initial condition. During the time evolution we observe that the vortex spectrum relaxes towards the equilibrium spectrum \eqref{eq:thermalKWspectrum}, displayed by the green dashed line. Note that the LIA prediction provides a reasonable prediction for scaling for small $k/k_0$-scaling, but as $k/k_0$ increases, it deviates from $1/\omega_k^{\rm GP}$ which reproduces well the data.

We have shown how a simple physical argument can predict the scaling of KW spectrum in equilibrium. However, in order to predict temporal dynamics, one needs to do further modelling. In the next section, we introduce a Langevin model that predicts the temporal behaviour of KWs close to equilibrium.

\subsection{Stochastic LIA equations for finite temperature quantum vortices} \label{sec:newNOISE}

In vortex filament models, finite temperature effects are typically taken into account using the Hall-Vinen Eq.~\eqref{Eq:VortexDyn}. In this framework, the LIA finite-temperature description is obtained by using the self-induced velocity ${\bf v}_{\rm sl}=-\frac{\Gamma \Lambda}{2\pi}{\bf s}'\times {\bf s}''$. It reads
\begin{equation} 
    \dot{{\bf s}}=-\frac{\Gamma \Lambda}{2\pi}{\bf s}'\times {\bf s}''+\alpha {\bf s'}\times({\bf v}_{\rm n}+\frac{\Gamma \Lambda}{2\pi}{\bf s}'\times {\bf s}'')-\alpha'{\bf s'}\times[{\bf s}'\times({\bf v}_{\rm n}+\frac{\Gamma \Lambda}{2\pi}{\bf s}'\times {\bf s}'')],\label{Eq:VortexDynLIA},
    \end{equation}
where we recall that $\alpha$ and $\alpha'$ are the mutual friction coefficients and ${\bf v}_{\rm n}=(v_{\rm n}^1,v_{\rm n}^2,v_{\rm n}^3)$ is the prescribed normal fluid. Note that in this model, thermal fluctuations are absent. 

We consider now the case of vortex line close to equilibrium at small temperatures, for which Kelvin wave amplitudes and the normal fluid are both small. Similarly to the $T=0$ case, the linear equations written in complex variables read
\begin{equation}\label{eq:LIA_HallVinen}
    \frac{\partial s}{\partial t}=-(\alpha +i(\alpha'-1))\frac{\Gamma \Lambda}{4\pi}\frac{\partial^2s}{\partial z^2}+ (i\alpha+\alpha')U
\end{equation}
where $U=v_{\rm n}^1+iv_{\rm n}^2$.

We now assume that thermal fluctuations are introduced in the system through the normal fluid. At low-temperatures, and within a classical approximation, the one time-statistics of the thermal bath is Gaussian and each Fourier mode of $U$ can be assumed to be independent. Furthermore, in reference \cite{Giuriato_2021} it was shown that the correlation time of thermal states is of order $\xi/c$, which is much smaller than typical large-scale KW periods. Under this assumption, it is natural then to assume that $U$ is a white noise in time. Rewriting Eq.~\eqref{eq:LIA_HallVinen} in Fourier space, and incorporating the previous considerations, leads to the following stochastic local induction approximation (SLIA) equation
 \begin{eqnarray}\label{eq:StochasticLIA}
     \frac{\partial \sk}{\partial t}&=&-\gamma k^2\sk+ \sigma\, \zeta(t)\\
     \langle \zeta(t)\rangle&=&0\\
     \langle \zeta^*(t)\zeta(t')\rangle&=&\delta(t-t'),
 \end{eqnarray}
where $\gamma=(\alpha +i(\alpha'-1))\frac{\Gamma \Lambda}{4\pi}$ and $\sigma^2= (\alpha^2+\alpha'^2)\langle\lvert U\lvert^2\rangle$. Note that $\gamma$ is a complex number and encompasses damping by mutual friction and oscillations due to KW dynamics. The amplitude of the noise is related to the normal fluid mean kinetic energy, which is proportional to temperature.

The complex Langevin Eq.~\eqref{eq:StochasticLIA} admits a simple solution in terms of the standard Wiener process $W_t$. It reads 
\begin{equation}
    \sk(t)=\sk(0)e^{-\gamma k^2 t}+\sigma \int_0^t e^{-\gamma k^2(t-s)}dW_s.
\end{equation}
The above formula allows to directly compute the correlation function of KWs by simple algebra and using basic properties of the Wiener process. The correlation function is given by
\begin{eqnarray}\label{eq:CorrKWs}
 \nonumber   \langle \sk(t)\sk^*(t')\rangle&=& \langle \sk(0)\sk^*(0)\rangle e^{-\gamma k^2t-\gamma^*k^2t'}+\sigma^2\int_0^t\int_0^{t'}e^{-\gamma k^2(t-s)-\gamma^*k^2(t'-s')}\langle dW_sdW_s'\rangle\\
    &=& \langle \sk(0)\sk^*(0)\rangle e^{-\gamma k^2t-\gamma^*k^2t'}+\sigma^2\int_0^te^{-\gamma k^2(t-s')-\gamma^*k^2(t'-s')}ds'\\
  \nonumber  &=& \langle \sk(0)\sk^*(0)\rangle e^{-\gamma k^2t-\gamma^*k^2t'}+\frac{\sigma^2 \left(e^{2Re[\gamma]k^2t-k^2 (t\gamma+t'\gamma^*)}- e^{-k^2(t\gamma+t'\gamma^*)} \right)}{2k^2 Re[\gamma]}.
\end{eqnarray}

Note that the equipartion spectrum \eqref{eq:thermalKWspectrum} follows from the previous expression by taking $t=t'\to\infty$
\begin{equation}
    \langle \lvert\sk\lvert^2\rangle=\frac{\sigma^2 4\pi}{2\alpha \Gamma \Lambda k^2}.
\end{equation}
The previous expression relates fluctuations ($\sigma$) and dissipation ($\alpha$) of KWs in equilibrium.

More interesting, expression \eqref{eq:CorrKWs} gives an explicit formula for the correlation function
\begin{equation}\label{eq:defKWCorrFunction}
    C_k(\tau)=\frac{Re[ \langle \sk(0)\sk^*(\tau)\rangle]}{ \langle \lvert\sk(0)\lvert^2\rangle}.
\end{equation}
Assuming that at $t=0$ all the KWs are in thermal equilibrium, formula \eqref{eq:CorrKWs} leads to
\begin{equation}
    C_k(\tau)=e^{-\alpha \frac{\Gamma \Lambda}{4\pi} k^2t }\cos{\left[(\alpha'-1) \frac{\Gamma \Lambda}{4\pi} k^2t\right]}.
\end{equation}
We expect then a decay rate given by the mutual friction coefficient $\alpha$ and a renormalisation of the KW frequency due to $\alpha'$.

Note that a more realistic description could be obtained by replacing the LIA dispersion relation $\omega_k^{\rm LIA}$ by a more accurate KW dispersion in the same spirit of predictions made in \cite{Giuriato2020How}.  With such a change, Eq.~\eqref{Eq:VortexDynLIA} becomes a non-local partial differential equation (with Laplacian replaced by a non-local operator), but with a simple representation in Fourier space and all the previous calculations trivially follow.

\subsection{Finite-temperature equilibrium Kelvin wave correlation function using the truncated Gross-Pitaevskii model}\label{subsec:FiniteTemp_KW_in_GP}

\begin{figure}[h]%
    \centering
    \includegraphics[width=1\textwidth]{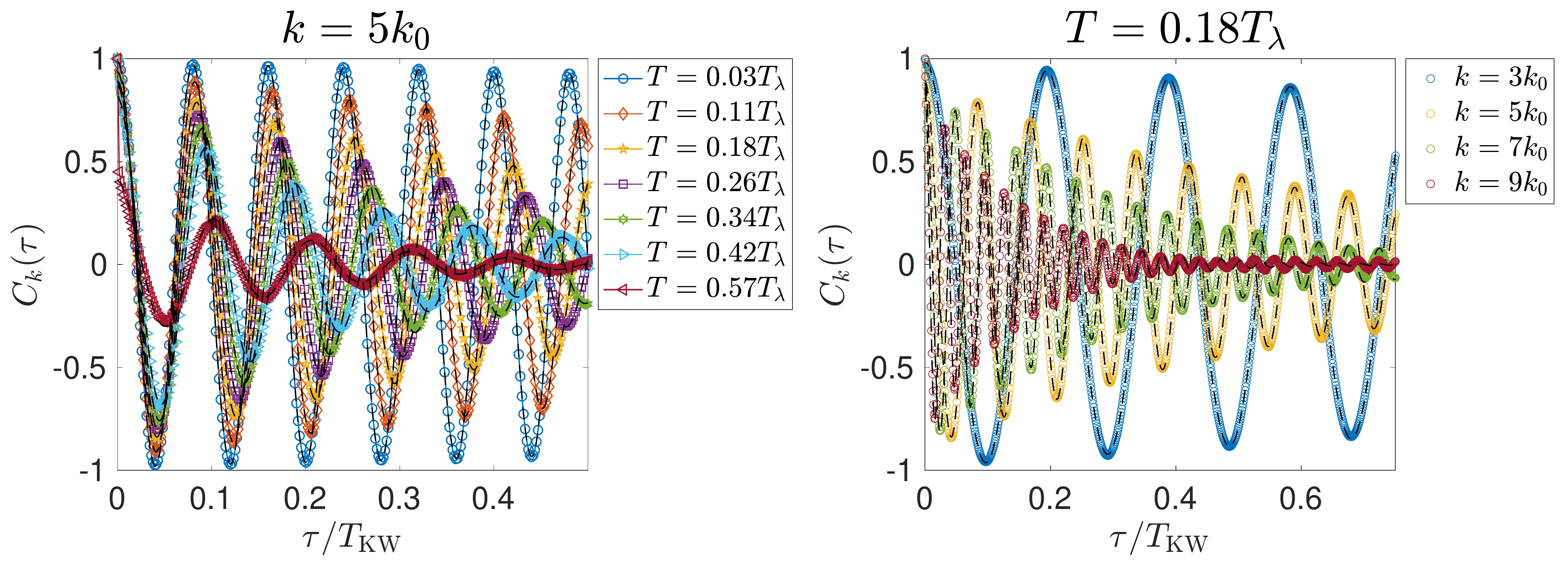}
    \caption{Equilibrium Kelvin wave correlation function \eqref{eq:defKWCorrFunction} obtained using the TGP equation. \textbf{left:} Correlation function at different temperatures for a fixed wave vector. \textbf{left:} Correlation function at different wave vectors for a fixed temperature. The dashed lines show the fit obtained using formula \eqref{eq:corrFit}.}\label{fig:corrfunction}
\end{figure}

We now study the correlation function \eqref{eq:defKWCorrFunction} using the TGP equation. We use the same numerical setting of Section \ref{subsec:eqTGPKW} and we compute the correlation function once the KWs have reached thermal equilibrium. The measured correlation functions for different temperatures and wave vectors are displayed in Fig.~\ref{fig:corrfunction}.
It is apparent from the figure that there is a clear temperature-dependent decay rate and a slowdown of the oscillation frequency. The quality of the data allow us to determine damping and frequency coefficients by fitting the temporal series with the fit
\begin{equation}\label{eq:corrFit}
    C_k^{\rm fit}(\tau)=A e^{-\lambda_k t}\cos{\Omega_k t}.
\end{equation}
The resulting fits are also displayed in Fig.~\ref{fig:corrfunction} in black dashed lines.

\begin{figure}[h!]%
    \centering
    \includegraphics[width=.85\textwidth]{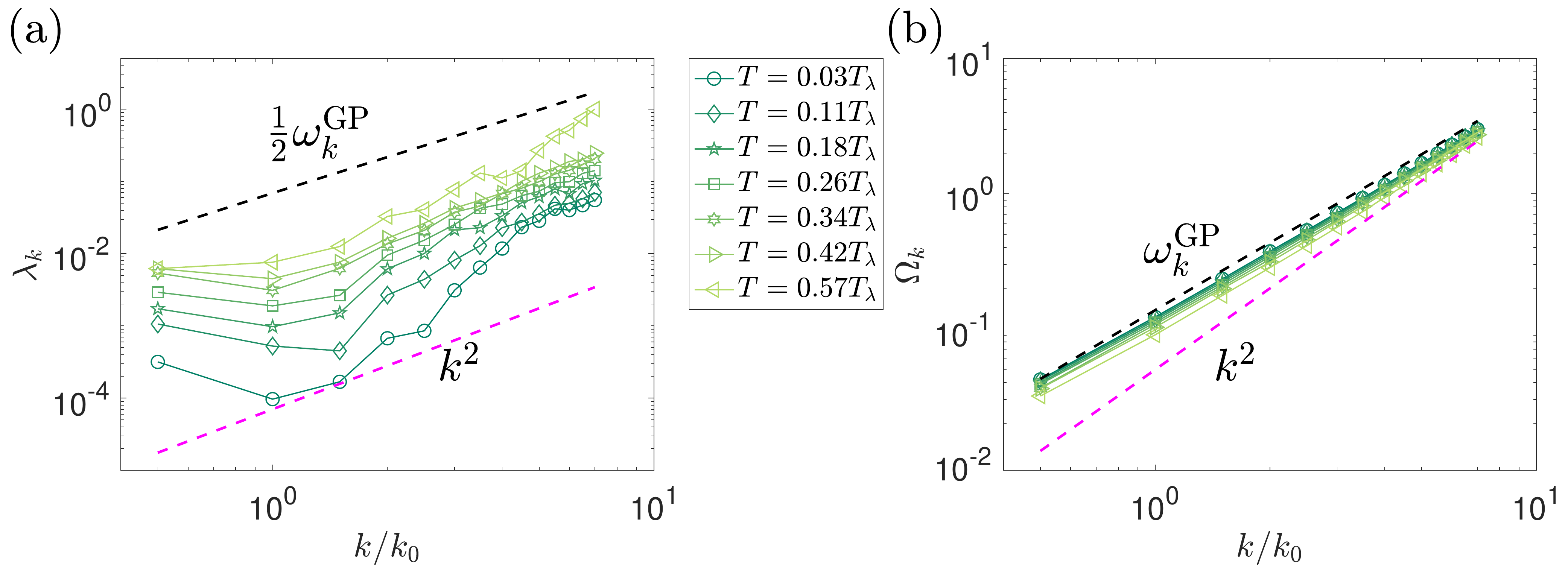}
    \includegraphics[width=.85\textwidth]{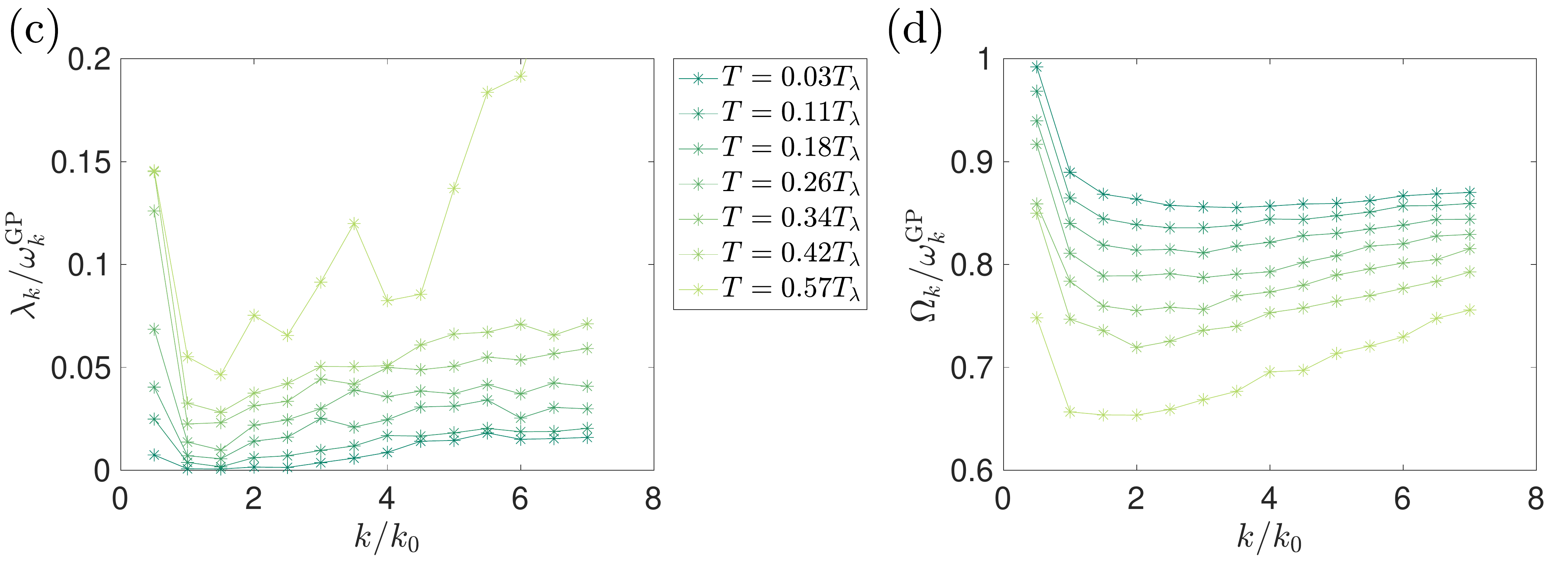}
    \caption{Equilibrium Kelvin correlation function \eqref{eq:defKWCorrFunction} obtained using the TGP equation. \textbf{left:} Correlation function at different temperatures for a fixed wave vector. \textbf{left:} Correlation function at different wave vectors for a fixed temperature. The dashed lines show the fit obtained using formula \eqref{eq:corrFit} (factor $1/2$ is arbitrary) and the LIA scaling $k^2$.}\label{fig:coeff}
\end{figure}

\begin{figure}[h]%
    \centering
    \includegraphics[width=1\textwidth]{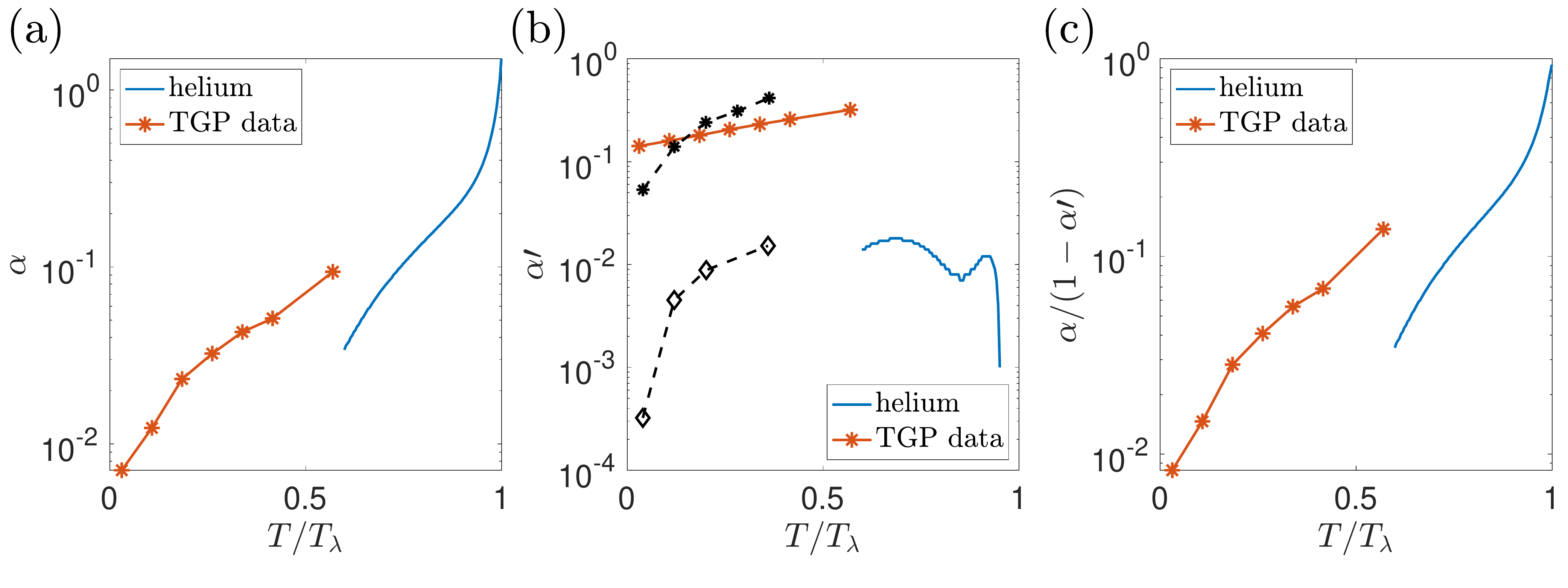}
    \caption{Measured mutual friction coefficients obtained with the TGP model. We also plot the measured helium mutual friction coefficient taken from \cite{HeliumProp_Donelly_Barenghi}. In figure \textbf{(b)}, we also show the measured TGP friction coefficients for the case of the vortex lattice and a vortex ring presented in Fig.~\ref{fig:mutualFriction} (same markers)}\label{fig:mutualFriction}
\end{figure}

In Fig.~\ref{fig:coeff} we present the results of the fit.
Figure  \ref{fig:coeff}.a displays the coefficient $\lambda_k$ for different temperatures. For comparison, we also plot the $k^2$-scaling expected from the LIA model and Kelvin wave dispersion relation for GP KWs given by Eq.~\eqref{eq:omegaKWGP}. A strong temperature dependence is observed, the higher the temperature, the larger the damping. Moreover, the scaling of $\lambda_k$ seems in agreement with the SLIA prediction \eqref{eq:defKWCorrFunction}. Similarly, Fig.~\ref{fig:coeff}.b displays the measured frequencies $\Omega_k$. There is a small temperature dependence producing a slowdown of the frequency. In addition, it is clear that frequencies do not follow the LIA scaling, but they agree without adjustable parameters with the KW dispersion relation for the GP model. In figures \ref{fig:coeff}.c-d we show $\lambda_k$ and $\Omega_k$ normalised by $\omega_k^{\rm GP}$. We then define the measured GP mutual friction coefficients by averaging over wave vectors $k/k_0\in(2,5)$ where curves are relatively flat. In addition, we average the ratio $\lambda_k/\Omega_k$, that from Eq.~\eqref{eq:defKWCorrFunction} should be independent of the KW dispersion relation and equal to $\alpha/(1-\alpha')$. Finally, the measured mutual friction coefficients are displayed in Fig.~\ref{fig:mutualFriction}.
For comparison, we also plot the measured helium mutual friction coefficient taken from reference \cite{HeliumProp_Donelly_Barenghi}. Helium and TGP mutual friction coefficient have been obtained at different temperature ranges. In the case of the TGP model, to perform the analysis carried out in this work at higher temperature is challenging, as thermal fluctuations become strong (including thermally excited vortices), which difficult the tracking of vortices.

The TGP-measured mutual coefficient $\alpha$ is remarkably of the same order of the one of helium, and it looks as a natural low-temperature continuation of the known (higher temperature) helium properties. On the contrary, the mutual friction coefficient $\alpha'$ seems not to agree with helium. Figure \ref{fig:mutualFriction}.b also display data form Fig.~\ref{Fig:alphap} (in black).

\section{Discussion and conclusions} \label{sec:Conclusion}

First, we briefly reviewed some of the mutual friction effects on vortex lines and rings that were obtained in references \cite{Krstulovic11} and \cite{Krstulovic11b}. In particular we reviewed the anomalous slowdown of rings that is produced by thermally excited Kelvin waves. Then, we studied the effect of mutual friction and thermal noise on the relaxation of Kelvin waves on straight vortex lines by comparing the results of full $3D$ direct simulations of the TGP with a simple new LIA stochastic model with mutual friction and thermal noise included. 

The new model allowed us to determine the mutual friction coefficient. By fitting the simple Langevin model prediction \eqref{eq:corrFit} to the TGP Kelvin wave correlation function shown in Fig. \ref{fig:coeff}, we have successfully reproduced the TGP correlation functions. 

The coefficients $\lambda_k$ and $\Omega_k$ obtained from the fit show a distinct temperature dependence, and the values of the measured mutual friction coefficient $\alpha$ is of a similar order of magnitude as those measured in superfluid helium. However, the value of the mutual friction coefficient $\alpha'$ obtained from the fit deviates from the standard measurements. It is rather puzzling and important to note that the values of $\alpha'$ obtained in the TGP context for the vortex ring (reviewed above in section~\ref{sec:mutual}) also differed significantly, suggesting that the geometry of the vortex configuration can affect the mutual friction coefficients. However, as there is no direct link between the two methods of determining $\alpha'$, this may just be a coincidence.

Finally, let us recall that to correctly describe superfluid liquid Helium, with a good equation of state and a dispersion relation with rotons, 
the GPE needs to be extended by including  non-local and higher order nonlinear terms \cite{Berloff_MotionsBoseCondensate_1999,Muller_CriticalVelocityVortex_2022,Muller_KolmogorovKelvinWave_2020}. 
It is an open problem that is left for further study to see if such a more quantitative description still lead to the same results: 
a 'good' $\alpha$ and a 'bad' $\alpha'$.

\section{Acknowledgments}
    G.K. was supported by the Agence Nationale de la Recherche through the project GIANTE ANR-18-CE30-0020-01 and acknowledges the support of the Simons Foundation Collaboration grant Wave Turbulence (Award ID 651471). M.E.B. acknowledges support from the French Agence Nationale de la Recherche (ANR QUTE-HPC Project No. ANR-18-CE46-0013).This work was granted access to the HPC resources of GENCI under the allocation 2019-A0072A11003 made by GENCI.
    Computations were also carried out at the Mésocentre SIGAMM hosted at the Observatoire de la Côte d'Azur.






\end{document}